# Attoampere Nanoelectrochemistry

Simon Grall[1], Ivan Alić[1], Eleonora Pavoni[2], Teruo Fujii[4], Stefan Müllegger[3], Marco Farina[2], Nicolas Clément[4*] and Georg Gramse[1,5*]

[1] Institute of Biophysics, Johannes Kepler University, Linz, Austria

[2] Department of Information Engineering, Marche Polytechnic University, Ancona, Italy

[3] Institute of Semiconductor and Solid-State Physics, Johannes Kepler University, Linz, Austria

[4] LIMMS/CNRS-IIS, Institute of Industrial Science, University of Tokyo, Tokyo, Japan

[5] Keysight Labs Austria, Keysight Technologies, Linz, Austria

* Correspondence: georg.gramse@jku.at; nclement@iis.u-tokyo.ac.jp

**Abstract**

Local electrochemical measurements and imaging at the nanoscale are crucial for the future development of molecular devices[1,2], sensors[3], materials engineering[4], electrophysiology[5,6] and various energy applications from artificial photosynthesis[7] to batteries[8]. The ultimate step towards single-molecule sensitivity requires the measurement of aA currents, which is three orders of magnitude below that of current state-of-the-art measurement abilities[9,10]. Here, we show electrochemical measurements at the sub aA level and <80 nm spatial resolution, that we reach by exploiting the ultra-high sensitivity of our GHz microscope for local faradaic interface capacitances. We demonstrate this for a well-known surface-bound ferrocene alkyl monolayer, a system that cannot be studied at the nanoscale unless large nanoarrays are used[11]. We report the simultaneous measurement of local cyclic voltammograms (CV) which provide atomistic information on the respective electron transfer reaction and reveal two molecular configurations with a similar redox energy potential - insights inaccessible by electrochemical ensemble measurements.

Pushing electrochemical research to the nanoscale would allow catalytic reactions to be followed at the molecular level, optimization of various devices made from nanoscale materials with electrochemical properties, and the study of electrophysiology with unprecedented resolution. Additionally, it would increase the understanding of molecular interactions in liquid, which is currently based on theoretical models of ensemble measurements. Towards these goals, scanning electrochemical microscopy, including the use of nanopipettes[12–14], has been developed to sense local electrochemical reactions. Moreover, scanning probes exploiting redox-cycling amplification to reduce the electrode dimensions while keeping a reasonable signal level[15–17] and electrochemical scanning tunnelling microscopy (EC-STM) have been used to study electrochemistry indirectly from a molecular electronics perspective[18]. While these techniques have enabled insights into the electrochemistry of single molecules, they remain applicable to only a few specialized research groups. Importantly, they are not sensitive enough to answer the open theoretical questions mentioned above or to obtain a signal at the nanoscale that is directly comparable to ensemble measurements. Therefore, many electrochemical systems remain unexplored at the nanoscale because the electrochemical signals are in the aA range, while current state-of-the-art measurements are in the fA range[2,11,17,19,20].

This study demonstrates how aA electrochemical measurements are achieved by measuring the faradic capacitive signal at a few GHz. While a current in the aA range cannot be measured directly due to an insufficient signal-to-noise ratio in pre-amplifiers, aF can be measured using high-frequency techniques, since the admittance of a capacitance $C$ scales with the frequency. Such an approach with aF resolution has recently been applied in the fields of semiconductor physics[21,22], molecular electronics[2] and biology[23].

The system selected in this study was a self-assembled monolayer (SAM) of ferrocene undecanethiol (FcC$_{11}$), which we probed with a modified STM that can deliver signals of several GHz to the tip. This method will have promising applications for high-frequency electronics through the THz gap and beyond, e.g., optical rectennas for solar cell applications without the Schockley–Queiser limitation[2,24]. Such systems have been extensively studied electrochemically but not at the nanoscale as they cannot benefit from redox-cycling amplification because the redox moieties are barely moving.

The electrochemical RF-STM combines a standard EC-STM setup with a high-frequency transmission line to the tip. Topography feedback is maintained by sensing the tunnelling current between the tip and the conducting substrate. The PtIr probe is coated with electrically insulating wax, exposing only the last 100 nm near the tip to the electrolyte, leading to a background current of <10 pA (Fig. S3).

The key to measuring currents three orders of magnitude below the state-of-the-art fA current amplifiers is the up-conversion to GHz frequencies. At these frequencies, the aA faradic current is converted into a capacitive impedance and integrated over each period. The $S_{11}$ signal scales linearly with frequency, $S_{11} \alpha \omega C_{farad}$ (Fig. 1), keeping the bandwidth of the measured capacitance signal. Sensing these extremely small faradic capacitance variations of a few molecules on top of the background capacitance of the cables and the upper part of the STM probe is technically challenging. We implemented a variable interferometric matching circuit that offsets the background impedance and solely amplifies the impedance variation sensed by the probe (Fig. 1). The network analyser measures the complex $S_{11}$ signal containing local information about the capacitance and conductance. With this matching network, we typically achieved a sensitivity of $\Delta C \approx 2$ aF at 2 GHz, which corresponds to $\Delta G \approx 25$ nS[25]. While increasing the detection frequency further would increase the sensitivity in the capacitance channel, the dielectric loss of water at these frequencies would annihilate this advantage.

To study electrochemical activity, we also implemented bipotentiostatic control of the substrate and probe, which enables stable STM feedback in the electrolyte for topographic imaging, local cyclic voltammetry and sensing the RF capacitance variation. While direct $S_{11}(V_{ec})$ cycles led to favourable signal-to-noise ratios, they were subjected to topography crosstalk, and electrochemical activity was difficult to distinguish from topographic variations. In order to increase further the sensitivity and selectivity for the electron transfer of the surface-bound electrochemical species, we applied an additional kHz modulation to the substrate. This signal modulated the electrochemical electron transfer, which was detected by a lock-in amplifier acquiring the complex derivative signal, $dS_{11}/dV$, corresponding to $dC/dV$ and $dG/dV$. All contributions that did not change with $V_{ec}$ were suppressed.

RF-STM allows for topography imaging and detection of electrochemical activity on the electrode surface. Fig. 2a shows the three-dimensional (3D) topography and local capacitance (colour scheme) of a $FcC_{11}$ SAM on gold at an electrochemical potential, $V_{ec}$, of 150 mV, close to its electrochemical redox potential. Moreover, the 3D colour maps show the conductance and capacitance recorded between $V_{ec}$ = 50 and 200 mV over the same image frame by sweeping $V_{ec}$ at 5 mV/s during the acquisition. For better clarity, Fig. 2b shows the histograms of consecutively recorded conductance and capacitance maps, exhibiting two peaks corresponding to the bright and dark areas in the maps. The peaks are separated by $\Delta G_{12}$ = 150 nS and $\Delta C_{12}$ = 20 aF, respectively. Although it is known that thiol monolayers may form different molecular organizations[26], our RF-STM results (Fig. 2) demonstrate the successful discrimination of subtle differences in electrochemical activity with unprecedented sensitivity and spatial resolution. Interestingly, a density difference of only 19±1 % (Fig. S5) between the two molecular organizations was obtained, a feature that ensemble measurements would miss due to similar peak positions. This finding is also in agreement with the theoretical prediction that electrostatic interactions between Fc molecules of the same molecular organization phase should negligibly affect the CV shape[11]. No other technique allows the resolution of such a small difference at the nanoscale.

As shown in Fig. 2c, no crosstalk between signals was exhibited. The lateral resolution of the microwave channels was estimated to be ~80 nm. With a capacitance noise value of 2 aF, the measured data were scaled as detailed in the methods section. To increase the accuracy, metrological calibration based on calibration kits[27] or retract curves[28] can be implemented in the future.

The advantage of the presented RF heterodyne capacitive detection of the oxidation state compared to classical direct current (DC) measurement is depicted in Fig. 1 and Fig. 3a. Under classical DC conditions, the acquired macroscale CV curves (Fig. 3b) result from the electrochemical charge transfer reaction, which oxidizes and reduces the surface-bound ferrocene molecules. The double bump observed in the CV is related to zones with larger roughness where electrostatic interactions between the Fc moieties are stronger. From Fig. 3b and Fig. S4, an average surface coverage of $\Gamma_T = 2.4\pm0.2$ Fc/nm² was determined in agreement with literature[2]. At the molecular level, the electrochemical charge transfer is based on the tunnelling of electrons from the gold electrode to the ferrocene molecules, accumulating charge ($Q_{ox}$) at the ferrocene-liquid interface, which attracts ions in the solution and induces the formation of an electrical double layer ($C_{dl}$). In the equivalent circuit model (Fig. 3a) this corresponds to a faradaic pseudocapacitance $C_{Far}(V_{ec})$ parallel to the geometric capacitance $C_{SAM}$ and in series with $C_{dl}$. Both $C_{Far}(V_{ec})$ and $C_{dl}(Q(V_{ec}))$ are intrinsically coupled by the charge state of the Fc/Fc$^+$ couple that follows a Fermi distribution centred on the oxidoreduction potential $V_0$ of the Fc/Fc$^+$ couple[11,29] such that

$$Q(V_{ec}) = \Gamma_T e_0 \left(1 + e^{-g\frac{(V_{ec}-V_0)F}{RT}}\right)^{-1} \qquad (1)$$

where is $e_0$ the elementary charge, $F$ is Faraday's constant, $R$ is the gas constant, $g \sim 0.67$ the broadening factor and $T$ is the temperature. At GHz frequencies well above the rate constant $k_{ET}$, the charges responsible for $C_{Far}$ are too slow to follow the electric field[30], and only $C_{dl} = dQ/dV$ is measured (Supplementary Note 3). Thus, the $\Delta C$ in a CV can be represented by first-order approximation as the derivative of the Fermi distribution, while $dC/dV$ is expected to follow the second derivative of the Fermi distribution (Fig. 3c, e). The electrochemical current at equilibrium simply corresponds to $\Delta C$ multiplied by the sweep rate (Fig. 3c). Of note, the CVs at the single-phase level are very similar to those of an ideal redox SAM at equilibrium and we estimate that the capacitance peak in the CVs would correspond to the charge transfer of Fc molecules in an area with a diameter of approximately 96±32 nm (Supplementary Note 5) and a signal sensitivity of 120 molecules/2 aF. This agrees well with the lateral resolution of 80 nm shown in Fig. 2. Therefore, for semi-quantitative analysis, the CVs obtained at a nanoscale by RF-STM can be directly compared to ensemble measurements, providing an extremely versatile method to acquire nanoelectrochemical images. Nevertheless, small variations from this simple model like the double bump only visible in the $dC/dV$ curve (Fig. 3e) and the $dG/dV$ signal (Fig. 3f) are still not addressed.

More detailed finite element models (FEMs) also account for the deviations from this simple approximation, which are related to the ionic media and the out-of-equilibrium effects that are the origin of the *dG/dV* signal (Fig. 3f). An important feature is that the RF conductance is very large compared to the DC tunnel conductance and does not show the characteristic exponential probe-sample distance dependency of a tunnelling process, but it changes as the *dC/dV*(*d*) curves only moderately with distance (Fig. S2). At a close distance ($d < 200$ nm), the *dC/dV*($V_{ec}$) curve for oxidation of the surface-bound ferrocene was measured (Fig. 3d). Meanwhile, only at $d > 200$ nm above the surface, the signal became insensitive to the electrochemical species.

The numerical FEM of the tip-sample system that captures these experimental observations is based on the Nernst–Planck-Poisson equations (as detailed in the methods section), and the potential distribution at the tip sample region both in the DC and GHz frequency regimes is shown together with an effective equivalent circuit model in Fig. 4a. Only the very end of the probe was exposed to the solution, while the upper part of the probe was isolated by a dielectric material. The DC potential completely dropped off in isolation, as expected; while at 1 GHz, the dielectric isolation was less effective but still led to a considerable potential drop, making the measurement more local. At the exposed part of the probe, only in the GHz regime, where the ions can hardly follow the field, the double layer capacitance started to relax. This agrees with the experimentally obtained *dC/dV* curve shown in Fig. 3d, obtained well above the surface, which follows the hyperbolic dependency of the diffusive layer capacitance with respect to the applied electrochemical potential. However, at a close tip sample distance, the relaxation time diminished again[31] (Supplementary Note 3).

Simulation of the spectra at a close distance (2 nm) above the SAM showed the experimentally observed peak in the conductance channel (compare Fig. 4b and 3f), and even the shoulder in the *dC/dV* signal is recovered (Fig. 3e), which could not be retrieved with the simple Fermi distribution. Therefore, the presence of ions in the solution and the diffusive double layer must be considered to obtain a quantitative agreement between the measured and calculated capacitance and conductance signals.

Of note, the charge transfer to the Fc is a function of time, reaction rate and voltage; thus, it is a dynamic process. Although the GHz modulation is faster than the reaction rate, appropriate selection of the kHz modulation frequency would allow for extraction of the dynamic properties of the molecular charge transfer, while the appropriate selection of the RF frequency and scan distance would allow the double layer dynamics to be investigated (Supplementary Note 3). Both topics will be further addressed in future research.

In conclusion, we have shown how electrochemical microwave microscopy locally resolves redox reactions of as few as 120 molecules, which corresponds to the transfer of the same number of elementary charges. To reach the required aA current sensitivity, we detected oxidation and reduction of surface bound redox species at microwave frequencies and amplified our sensitivity dramatically while maintaining the bandwidth. At these frequency ranges that are well above the electrochemical rate constants and diffusion times, we sense the change of the faradic capacitance instead of the current. Our technique sets the groundwork for a wide range of applications, including electrochemical catalysis, charge transport in molecular biophysics, and studying quantum mechanical resonance effects in complex molecules, which were completely inaccessible until now.

**Methods**

*Electrochemical setup*

The experimental setup is described in Fig. 1. It consists of a RF-STM microscope and a potentiostat connected to a Pt counter electrode (CE), an Ag/AgCl reference electrode (RE) and as working electrodes a gold(111)-on-mica substrate (Phasis, Switzerland) (WE) and the RF-STM microscope tip (PtIr, 80/20). A self-assembled monolayer (SAM) of ferrocene undecanethiol ($FcC_{11}$) was grafted on the gold substrates by incubation in 1 mM $FcC_{11}$ (Sigma Aldrich) in ethanol for at least 12 h in closed vials in ambient conditions. The gold-on-mica samples were only blue-flamed before the incubation, and rinsed with ethanol after incubation. Since a STM-based measurement setup is used, it is not desirable to have a densely packed monolayer that may hinder the tunnelling with reported thicknesses of 1.8 nm for such $FcC_{11}$ SAMs[32]. The samples are placed in a homemade electrochemical cell. Bottom contact for bias and electrochemical potential sweeping is made using silver paste. The electrochemistry is done in 1 mM $NaClO_4$ water solution bubbled with nitrogen immediately before every experiments. All parts of the setup are controlled through the microscope's controller and software (Keysight).

*RF setup*

The RF-STM consists of a 5400 Keysight AFM mount and a STM scanner modified for RF modulation at the tip. A bias tee separates low frequency currents (tunnelling and DC faradic currents) and the RF signal. A vector network analyser (VNA) (Keysight) serves as the RF source. The RF transmission line is described Fig. 1 and consists of an interferometric setup comprising a voltage-controlled attenuator, a directional coupler and a hybrid coupler. An AC voltage of 50 mV is applied to the sample with a frequency ranging from 30 kHz to 80 kHz depending on the experiment. The reflected signal ($S_{11}$) is obtained from the VNA and the derivative of $S_{11}$ versus the AC voltage applied to the sample ($dS_{11}/dV$) is obtained after direct down-modulation of the outgoing RF signal in a mixer. Therefore, the outgoing RF signal is multiplied with the back-reflected signal, which has been amplified by a low noise amplifier. This down modulated signal is fed into a lock-in amplifier giving the amplitude and phase of $dS_{11}/dV$. By introducing this additional low frequency (kHz) electrical modulation between the substrate and the probe, we remove the drift and dramatically improve the sensitivity to the electrochemical reaction.

*Data scaling*

In order to map complex $S_{11}$ data to the complex admittance plane $Y$, containing a conductance channel ($\Delta G$) and a capacitances channel ($\Delta C$), we can assume small variations in $S_{11}$ and use the first-order approximation relationship:

$$\Delta S_{11} = ae^{i\phi}\Delta Y = ae^{i\phi}(\Delta G + i\omega\Delta C)$$

While *a* defines the scaling of the admittance with respect to the measured $S_{11}$ signal, $\phi$ defines the rotation in the complex plane. The scaling factor *a* can be obtained by assessing the noise obtained on $S_{11}$ data, and comparing it with the noise floor obtained on independent measurements in dry environment. In our measurement frequency we found the noise floor to be around 2 aF for capacitances measured using such technique[28,33]. Being aware that the data presented here are obtained from the integration of $dS_{11}/dV$, a similar estimation of $G$ from $S_{11}$ data was done and led to results similar to the $G$ obtained out of $dS_{11}/dV$ (Fig. S1). Though $S_{11}$ signals suffer from significantly more crosstalk with topography and barely exhibits a capacitance signal (which is expected here because of stray capacitances), this confirms that the estimation of the capacitance and conductance from the noise on the integrated values of $dS_{11}/dV$ is not arbitrary.

*Finite Element Modelling*

FEM was carried out with COMSOL Multiphysics 5.5 (2D axisymmetric, Nernst-Plank-Poisson Equations, statics, time domain, frequency domain perturbation). The simulation geometry for quantification resembles the experimental conditions and is shown in Fig. 4a. The model consists of a 5 µm high STM probe modelled as a truncated cone with a cone angle of 15°. The tip has a spherical apex with a radius of 100 nm and is located at a distance $d$ above the sample. The SAM is modelled as a 1.5 nm thick dielectric layer with a dielectric permittivity of 2. The voltage dependent surface charge on the SAM is calculated according to eq. 1 using the Boundary ODE interface. To achieve accurate results and convergence of the solution, meshing was set to 0.1 nm on the exposed probe surface and the SAM using the boundary layer mesh. For the time dependent solution probe and substrate were set to apply the voltage sweep identical to the experiment and the upper part of the simulation box was set to ground. Concentration on the upper part of the simulation box was set to 10 mM. For the frequency perturbation step the Terminal was set to the probe and the ground was set to the substrate. With this model the $C(V)$ and $G(V)$ voltage sweeps were calculated and differentiated.

## Acknowledgements

The authors acknowledge funding from the ATTRACT project funded by the EC (grant agreement No 777222), the NMBP project Nanobat (grant agreement No 861962), the Austrian FWF Project P28018-B27, and the European Research Council (ERC) under the European Union's Horizon 2020 research and innovation program (grant agreement No 771193). The authors would also like to thank Dominik Farka, Nicolas Lesniewska, who helped greatly on the preliminary experimental setup, and Dr. David Toth, for the invaluable inputs and discussions during experimental processes.

## Author Contributions

SG performed the experiments, analysed the data, wrote the paper. IA and EP contributed materials/analysis tools. SM and MF contributed materials/analysis tools, wrote the paper. NC conceived the experiments, analysed the data, wrote the paper. GG conceived and designed the experiments, performed the experiments, analysed the data, wrote the paper.

## Competing Interests statement

The authors declare no competing interests.

**Figures**

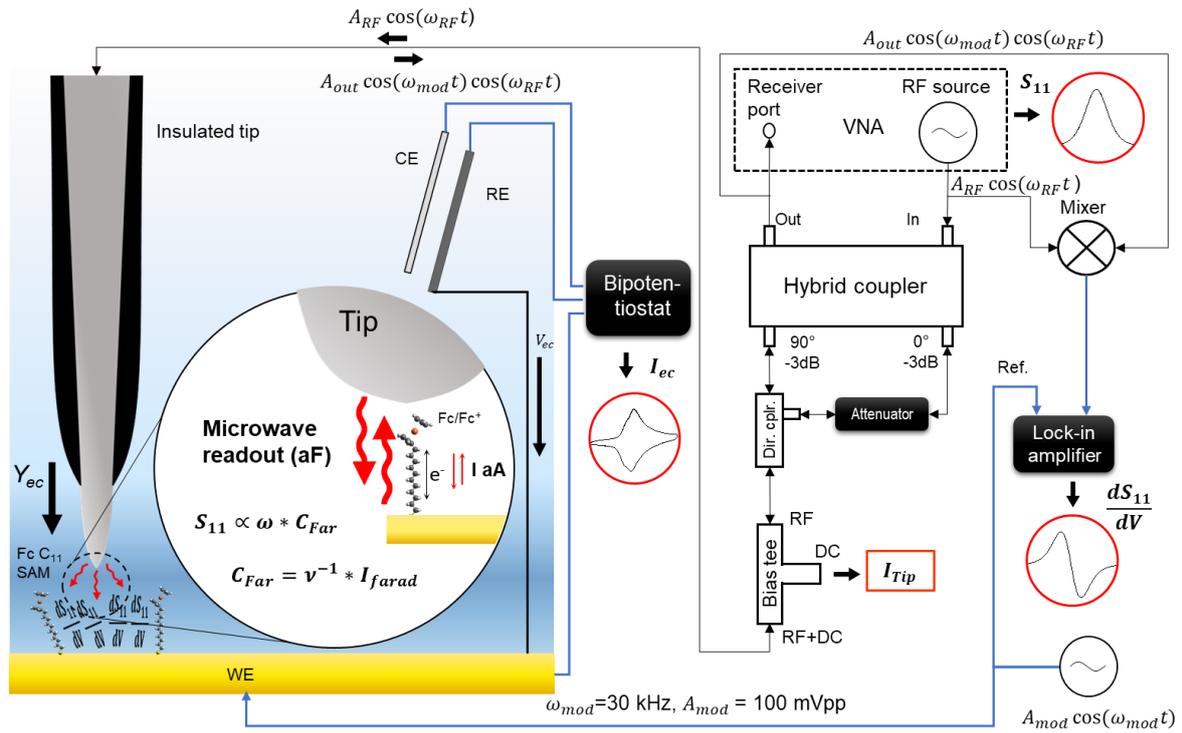

**Fig. 1 | Schematics of the radio frequency-scanning tunnelling microscope (RF-STM) used in this work.** The left side of the figure shows the electrochemical part of the setup, with a bipotentiostatic organisation with a PtIr tip and a gold sample as working electrodes, a Pt counter electrode and an Ag/AgCl reference electrode. A bias of 100 mV is maintained between the tip and the gold sample. The right side of the figure shows the transmission line of the RF signal, with a vector network analyser (VNA) as the RF source ($A_{RF} \cos \omega_{RF}$) and receiver for the $S_{11}$ signal. An interferometer configuration made of a hybrid coupler, a directional coupler and an attenuator is used to amplify the variations sensed at the tip-sample interface. A voltage sweep with a sweep rate of $v$ and an alternating voltage ($A_{mod} \cos \omega_{mod}$) are sent to the sample, which modulates the amplitude of the electrochemical electron transfer between the ferrocene head and the gold sample ($A_{out} \cos \omega_{mod} \cos \omega_{RF}$). This signal is directly down-modulated in a mixer and acquired by a lock-in amplifier, giving the $dS_{11}/dV$ signal. A bias tee separates the RF signal from the low-frequency tunnelling and electrochemical faradic signals.

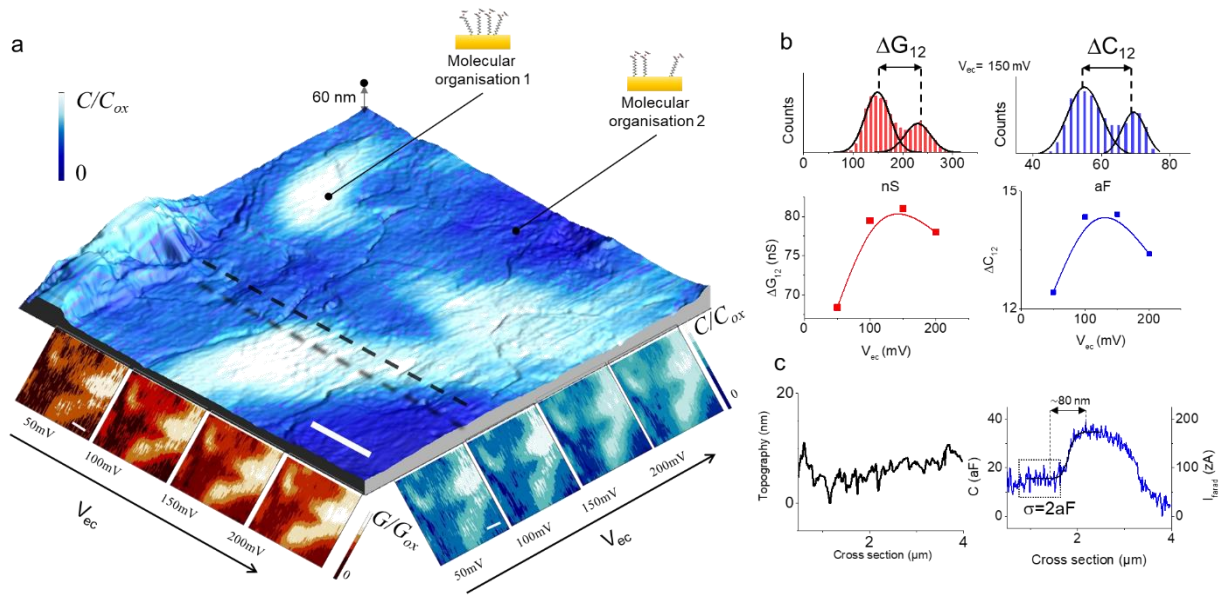

**Fig. 2 | Nanoscale molecular organisation. a** Three-dimensional (3D) representation of a self-assembled monolayer of FcC$_{11}$ grafted on gold at an electrochemical potential ($V_{ec}$) of 150 mV. The voltage was swept at $v$ = 5 mV/s on the sample during this experiment. The colour overlay represents the capacitance $C$ (at $V_{ec}$ = 150 mV), and the colour maps below the 3D image represent the capacitance and conductance ($G$) on the same sample between 0 mV and 200 mV. Scale bar is 800nm. **b** Histograms of the $C$ and $G$ colour maps at $V_{ec}$ = 150 mV. The two peaks are distinguishable and correspond to the two areas visible on the capacitance overlay in **a**. The peak distance ($\Delta G_{12}$ and $\Delta C_{12}$) is plotted versus the electrochemical potential $V_{ec}$ for both histograms. The solid line on the $\Delta G_{12}$ and $\Delta C_{12}$ plot is a guide to the eye. **c** Cross-section corresponding to the dashed line in **a** showing the topography and the capacitance channels. No crosstalk is observed, and a lateral resolution of approximately 80 nm is obtained for the capacitance channel. The capacitance is converted into current knowing the sweep rate (ν = 5mV/s), which gives a current on the order of 150 zA.

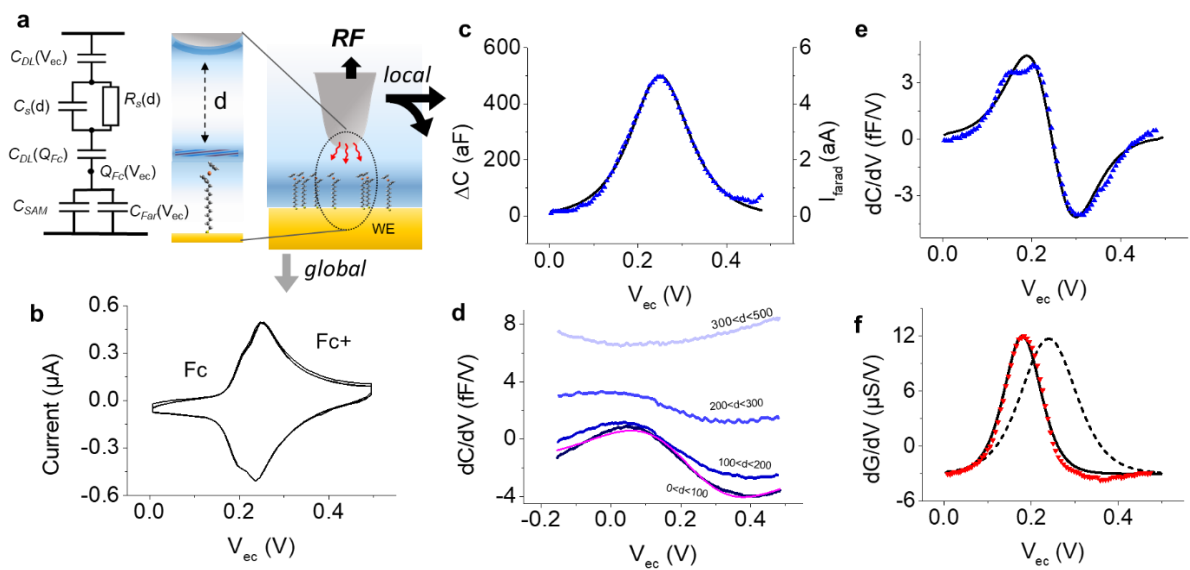

**Fig. 3 | Direct current (DC) and radiofrequency (RF) cyclic voltammograms. a** Schematic depicting the conditions in which the DC and RF signals are obtained. The RF-scanning tunnelling microscope tip was placed over a self-assembled monolayer of FcC$_{11}$ grafted on gold while the sample voltage ($V_{ec}$) was swept at $v$ = 10 mV/s. The corresponding equivalent circuit is shown on the left, emphasizing the presence of two double layers and its dependency on the tip-surface distance (*d*) and the sample pseudocapacitance C$_{far}$($V_{ec}$). **b** The DC cyclic voltammogram acquired by the potentiostat at the gold electrode. **c, e** Cyclic voltammograms of Δ*C* and *dC/dV*, with black lines representing the fits using the Fermi distribution first (**c**) and second derivatives (**e**). **d** *dC/dV* acquired at different distances during the potential sweep. **f** The *dG/dV* channel during the same experiment depicted in **c, e**. The solid and dashed lines represent the Fermi distribution first-derivative fits, with the solid line fitting the *dG/dV* datapoints and the dashed line showing where the datapoints would be expected if centred on the same potential as in **c, e**.

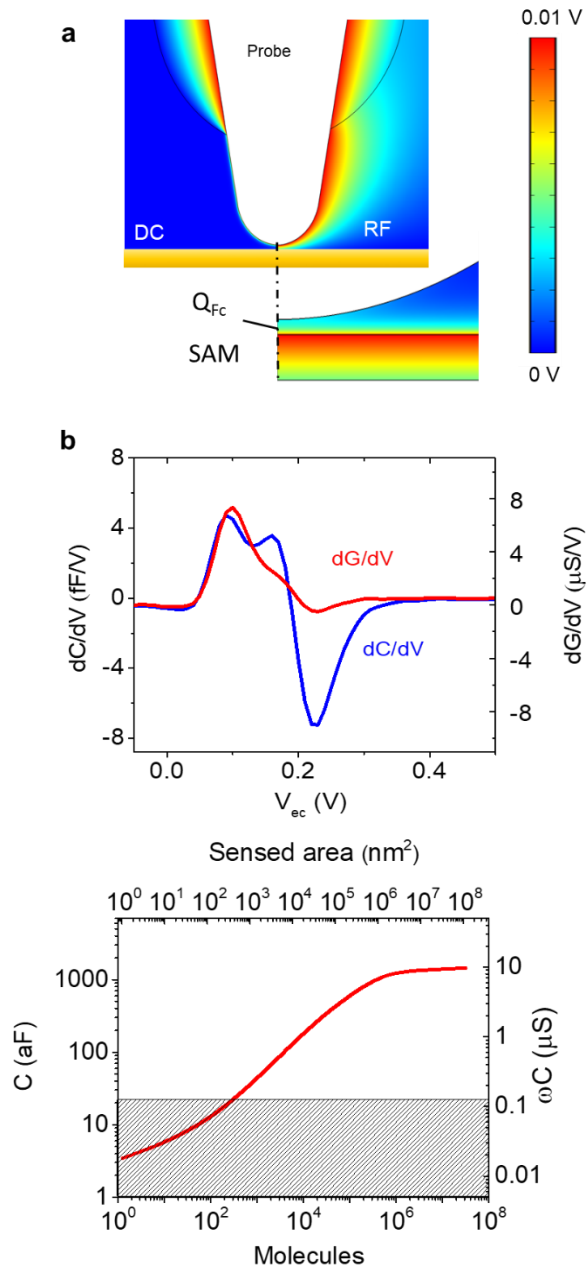

**Fig. 4 Advanced modelling and sensitivity analysis from finite element simulations. a** Finite element simulation of the insulated tip over the self-assembled monolayer (SAM) in the direct current and radiofrequency regimes. The zoom-in below shows the potential distribution in the SAM. **b** Simulated *dG/dV* and *dC/dV* curves with the probe at $d = 2$ nm and the experimental conditions as in Fig. 3c. **c** Sensitivity analysis of the capacitance signal. The expected capacitance signal is dependent on the FcC$_{11}$ area, and the number of molecules is calculated. The dashed area corresponds to the sensitivity limit of the setup.